\documentclass[aps, 12pt, amsmath, amssymb, noshowpacs, preprint,preprintnumbers,prd]{revtex4}  
\usepackage{amsmath}    
\usepackage{amsfonts}  
\usepackage{epsfig}  
\usepackage{subfig}  
\newcommand{\beq}{\begin{equation}}  
\newcommand{\eeq}{\end{equation}}  
  
\newcommand{\beqn}{\begin{eqnarray}}  
\newcommand{\eeqn}{\end{eqnarray}}  
  
\newcommand{\bP}{\begin{Parallel}{120mm}{80mm}}  
\newcommand{\eP}{\end{Parallel}}

\newcommand{\p}{\partial}  
\newcommand{\ph}{\varphi}  
  
\newcommand{\la}{\langle}  
\newcommand{\ra}{\rangle}

\newcommand{\rar}{\rightarrow}      

\begin{document}      

\preprint{ITEP-TH-29/11}    

\title{ Baryon as dyonic instanton}    

\author{A. Gorsky}    

\author{A. Krikun}        

\affiliation{ITEP, B. Cheryomushkinskaya 25, 117218 Moscow, Russia}    

\begin{abstract}  
We discuss  the baryon in the holographic QCD framework and focus on the role of the bifundamental scalar field, realizing chiral symmetry breaking.   We suggest the interpretation of  a baryon as a  dyonic instanton within the Atiyah-Manton-like  approach  for the flavor gauge group    in the peculiar cylindrical ansatz in five-dimensional theory.  Our approach provides  the new mechanism of the stabilization of the baryon size.    
\end{abstract}      

\maketitle        

\section{Introduction}    

The topological interpretation of the baryon has been found  long time ago in the Skyrme model \cite{witten1}. Later it was  recognized that the holography provides  very transparent  geometrical realization of  topological nature of the baryon.  The baryonic vertex has been found in \cite{witten2} and its  realization as the instanton in the flavor gauge group has been clarified  in \cite{son}. The relation between the Skyrmions and the instantons  in five dimensions has been noticed in  \cite{am} where the  Skyrme field gets interpreted as a kind of  monodromy in the instanton  background. The Atiyah-Manton interpretation has been realized  dynamically in the model with the domain wall localized in the fifth  dimension \cite{tongjap}. 

The interpretation of the baryonic charge as the instanton charge in D=5 \cite{Sakai} or difference of the Chern numbers \cite{WulPom3}  works perfectly however some  characteristics of the nucleon  are harder to obtain. In particular the nucleon mass is determined by the instanton action   and the mechanism of stabilization of the instanton size is required.  In the conventional Skyrme model  the baryon size is stabilized by the higher derivative terms and  the holographic mechanisms were discussed in the literature as well \cite{Sakai, WulPom1, WulPom2}. However there  is no satisfactory size stabilization pattern which would be free from drawbacks. There also exist the other approach to address the baryon physics in holographic models by introducing the fermion field in the bulk \cite{fermi_baryon}, but it is not related to the present study, because it doesn't account  for the size of the baryon at all.      

On the other hand  there is   the mysterious Ioffe's  formula  \cite{ioffe} derived via QCD sum rules
\beq
\label{Ioffe}
m_N^3= -8\pi^2 <0|\bar{q}q|0>.
\eeq
Its status was questionable for decades since there was no alternative  derivation of this suggestive relation. It certainly implies that the  nucleon size and mass are fixed by some stabilization mechanism  involving the chiral symmetry breaking.

Motivated by these unsolved problems we shall look for  another  solution to the equations of motion in D=5 holographic dual with the proper quantum numbers.  The proper candidate is the dyonic instanton which generalizes  the conventional instanton in the D=5 theory with the scalar field  in the adjoint representation. Such solution has been found  in \cite{lt} and  it carries the additional charge  quantum number as well as the  angular momentum. The key point  is that the instanton  size  is fixed by the vacuum expectation value (VEV) of  the scalar $v$ and the electric charge $Q_e$ of solution.  
\beq 
\label{rho}  
\rho = \sqrt{\frac{Q_e}{4 \pi^2 v}}  
\eeq  
If the Chern-Simons term is added to the 5D action the moduli  space of the dyonic instanton gets modified \cite{cs}.        

We shall use such solution in the holographic QCD framework  when the gauge group $SU_L(N_F)\times  SU_R(N_F)$ on the flavor  branes is broken  to the diagonal one by the chiral condensate \cite{Erlich}.  It is convenient to describe the condensate via the scalar  field in the bifundamental representation with respect to the  bare gauge group. Upon the chiral symmetry breaking we find  ourselves in the situation where the dyonic instantons could be look  for.

Comparing to the Lambert-Tong solution \cite{lt} there are some  essential differences. First, instead of the flat space we should  start with the instanton in $AdS_4$ geometry.     Secondly our dual theory involves the scalar in the bifundamental  representation hence we need to clarify the physical meaning  of the corresponding charge and VEV which would stabilize the  dyonic instanton size. We shall work in the chiral limit hence  according to the holographic dictionary the boundary behavior in the radial coordinate of the VEV of the bifundamental yields the chiral condensate.  On the other hand the scalar  interacts with the axial gauge field hence  the proper second charge of the dyonic instanton is identified with the  axial charge in the physical space. This identification  fits with the   non-vanishing axial charge of the nucleon determined by the matrix  element of the axial current   
\beq  
\label{axial_charge}  
<N|A|N> \sim g_A  
\eeq       

We shall look for the dyonic instanton type solution in    the simplest holographic hard-wall  model \cite{Erlich} in  convenient cylindrical ansatz \cite{Witten_inst} which describes   the $SO(3)$ symmetric solution in the physical space. It will be found that the  Atiyah-Manton type approach \cite{am} can be generalized in the theory  with  an additional scalar field. The domain wall type solution  in $AdS_4$ geometry is described and  its second  charge   (similar to the electric charge of dyonic instanton) is identified with   the axial charge of the baryon. Hence we obtain the new mechanism  for the baryon size stabilization where both the chiral condensate and  the baryon axial charge play the crucial role. This qualitatively fits   with the Ioffe's formula for the baryon mass.        

The paper is organized as follows. In Section 2 we remind  the Lambert-Tong solution and its key features. In Section 3  we present the generalization of dyonic instanton solution  for the holographic QCD. Section 4 is devoted to the comments  on the brane interpretation of the solution found. Summary of our findings   and the open  questions can be found in the Conclusion.     

\section{On Dyonic instantons }          

The dyonic instanton \cite{lt}  is solution to the conventional $D=5$ equation of motion  
\beq  
F_{\mu\nu}= *F_{\mu\nu},\qquad D_{\mu}\phi =E_{\mu},\qquad D_{0}\phi =0, 
\eeq  
where $\phi$ is the scalar field and $E_{\mu}$ is electric  field in $R^4$. The BPS  formula yields for its mass  
\beq  
M= \frac{4\pi ^2}{g^2} |I| + |vQ_e|,  
\eeq
where $|I|$ is the instanton charge and $v$ is the VEV  of the scalar. The BPS-ness is provided  by the combined effect of  the scalar,  electric field  and   the "running" of the instantons.

There are dyonic instantons with the different symmetries  with respect to the space rotations. In what follows we shall   discuss the solution with the spherical symmetry  however the most transparent  physical picture is realized by the tubular  D2 brane stretched  between two parallel D4 branes \cite{dyonic2} which we shall briefly  describe in this Section. To identify all charges  it is useful to discuss the  worldvolume Lagrangian of the tubular D2 brane of constant radius  $R$ in flat Euclidean space-time \cite{dyonic2}  
\beq  
L= - \sqrt{R^2(1-E^2) +B^2},  
\eeq  
where $E,B$ are the worldvolume electric and magnetic fields.  The corresponding canonical momentum reads as  
\beq  
\Pi =\frac{\partial L}{\partial E}= \frac{R^2 E}{\sqrt{R^2(1-E^2) +B^2}}  
\eeq  
and Hamiltonian density is  
\beq 
{\cal H}= R^{-1}\sqrt{(\Pi^2 +R^2)(B^2 + R^2)}  
\eeq  
There are two quantum numbers  
\beq  
Q_F=\frac{1}{2\pi}\oint d\phi \Pi , \qquad Q_0=\frac{1}{2\pi}\oint d\phi B  
\eeq  
corresponding to the F1 and D0 conserved charges per unit length carried  by the D2 tube. 
The tension  reads as  
\beq  
T=\frac{1}{2\pi}\oint d\phi {\cal H},  
\eeq  
which equals at the solution to the equation of motion to  
\beq  
T=|Q_F| + |Q_0|  
\eeq  
Therefore  the D2 brane tension does not contribute  and the cross section of the tube behaves as the "tensionless" object  which explains the arbitrary form of the tube cross section.    

The solution carries  crossed electric and magnetic fields  that is the Poynting  vector does not vanish and yields  
\beq  
M_{\phi}=\Pi B  
\eeq  
Therefore there  is non-vanishing angular momentum of  dyonic instanton per unit length \cite{town}  
\beq  
L= Q_F Q_0  
\eeq  
directed along the axis of the cylinder.  One can calculate it  exactly at the cross-section  of the supertube \cite{town}  
\beq  
L=\oint d s \left(x_3\frac{\partial x_4}{\partial s} -  x_4\frac{\partial x_3}{\partial s} \right)  
\eeq  
It is this angular momentum which provides the stabilization of the  radius of the tubular D2 brane stretched between two D4 branes.      

The conventional dyonic instanton solution involves the  physical time hence upon the analytic continuation  to the Minkowski space it corresponds to some tunneling process.  The key point concerns the existence of the negative mode  in the fluctuation spectrum at the top of the solution  which would provide a sort of instability  in the Minkowski space. Such negative mode responsible  for the expanding of the radius of the solution has been  found for the large radii in \cite{emparan}.  Hence at least at large radius of the dyonic instanton  fixed by its angular momentum there is  negative  mode in $R^4$ which supports the bounce interpretation.  However at small values of the angular momentum there  are no negative modes and the solution behaves as instanton.            

\section{Dyonic instanton in holographic QCD}    

To realize the dyonic instanton type solution in holographic framework we take the simplest ``hard-wall'' model of AdS/QCD \cite{Erlich}. It posses two gauge fields, corresponding to the QCD flavor group of $SU(2)_L \times SU(2)_R$ and a bifundamental scalar $X$, responsible for the chiral symmetry breaking (it is actually crucial for construction of dyonic instanton \cite{lt}). The action of the model is simple  \begin{equation}
\label{action}  
S = \int d^3 x \ dt \ dz \left\{ \frac{1}{z} \left( - \frac{1}{4 g_5^2} \right) (F_L^2 + F_R^2) + \frac{\Lambda^2}{z^3} (D X)^2 + \frac{\Lambda^2}{z^5} \  3|X|^2 \right\},  
\end{equation}  
where $D\cdot = \p \cdot - iL\cdot + i \ \cdot R$ is an appropriate covariant derivative, $g_5$ and $\Lambda$ are the 5D coupling constant and normalization constant of scalar field, which are fixed by the comparison of 2-point correlation functions of vector and scalar currents, computed in the model and in QCD sum rules \cite{Erlich, Krik1, Krik2}. The model is embedded in the AdS space with a hard wall, placed at certain radial coordinate:  
\begin{equation}  
ds^2 = \frac{1}{z^2} (-dz^2 -dx_i^2 + dt^2), \qquad z<z_m  
\end{equation}  
As was shown in \cite{Sakai, witten2}, generally the Skyrmion in holographic model is represented by the topologically charged 4D field configuration in $t=const$ slice of $AdS_5$, and the baryon number identified with  the topological charge of this ``$AdS_4$ instanton''. For the particular model with two gauge fields the baryon number was defined in \cite{WulPom1,WulPom2,WulPom3}.  
\begin{equation}  
\label{baryon}  Q_B = \frac{1}{32 \pi^2} \int d^3 x \int \limits_\epsilon^{z_m} dz \ \left \la F_L \tilde{F_L} - F_R \tilde{F_R} \right \ra,  
\end{equation}  
where $\tilde{F}$ is a dual field strength tensor and angle brackets denote the trace over flavor (gauge group) indices.    

In what follows we will describe a classical solution in the model (\ref{action}) with finite action and nonzero charge (\ref{baryon}), incorporating nontrivial dynamics of the scalar field $X$, which can be named ``holographic dyonic instanton''.  The similar problem was considered in \cite{Wulzer4} and we will point out the difference later. First of all we will take the ``cylindrical'' ansatz, proposed in \cite{Witten_inst} and used in holographic context in \cite{WulPom1,WulPom2,WulPom3, Wulzer4}: 
\begin{align}  
\label{ansatz}  
A_j^a &= \frac{1 + \ph_2 (r,z)}{r}  \epsilon_{jak} \frac{x_k}{r} + \frac{\ph_1  (r,z)}{r} \left( \delta_{ja} - \frac{x_j x_a}{r^2} \right) + A_r(r,z) \frac{x_j x_a}{r^2} \\ 
A_z^a &= A_2(r,z) \frac{x_a}{r}  
\end{align}  
for $A^L$ and $A^R$, and  
\begin{equation}  
X = \chi_0(r,z) \frac{\mathbf{1}}{2} + i \chi_1(r,z) \frac{\tau^a x^a}{r}  
\end{equation}  
for bifundamental scalar field. Here $r^2 = x_1^2+x_2^2+x_3^2$ is a spacial radial coordinate, $\tau^a$ -- are the $SU(2)$ group generators, $a,b,\dots = 1\dots3$ are the group indices and $i,j,\dots = 1\dots3$ -- the spacial ones. Moreover, the P-parity conditions (see \cite{WulPom2})  
\begin{equation}  
A^L_i (x,z) = -A^R_i (-x,z), \qquad L_z(x,z) = R_z(-x,z)  
\end{equation}  
force additional constraints on the new fields  
\begin{equation}  
\ph_1^L = - \ph_1^R, \qquad  \ph_2^L = \ph_2^R, \qquad  A_r^L = -A_r^R, \qquad A_z^L = -A_z^R, 
 \end{equation}  
and we will use only left fields below, omitting the index $L$.  This ansatz reduces our model to the 2-dimensional Abelian gauge model with two charged scalars  
\begin{align} 
\boldsymbol{\ph}  = \ph_1 + i \ph_2 &\equiv \ph \ e^{i \alpha}, \\
\notag  
\boldsymbol{\chi} = \chi_0 + i \chi_1 &\equiv \chi \ e^{i \beta},  
\end{align}  
with the action  
\begin{align}  
\label{2D-action}  
S = 4 \pi \int dt \int  dr dz \Bigg\{-  \frac{1}{2 g_5^2}  \Big[ & \frac{2}{z} |D_a \boldsymbol{\ph}|^2 + \frac{1}{2} (F_{ab})^2 + \frac{1}{r^4} (1 - |\boldsymbol{\ph}|^2)^2 \Big]  \\
\notag  
- \frac{\Lambda^2 r^2}{z^3} \Big[  & \frac{1}{2} |D_a \boldsymbol{\chi}|^2 + \frac{1}{2 r^2}  (|\boldsymbol{\chi}|^2 + |\boldsymbol{\ph}|^2 - |\boldsymbol{\ph} + \boldsymbol{\chi}|^2) \Big] \\
\notag
+ \frac{\Lambda^2 r^2}{z^5} \ & \frac{3}{2} |\boldsymbol{\chi}|^2  \Bigg\},  
\end{align}  
where $a,b =(z,r)$, $D_a = \p_a + i A_a$ and $F_{ab} = \p_a A_b - \p_b A_a$.  Apart of the holographic mass term for the $X$ field, there are two interesting potential terms in the action. The third term in the first line is the usual potential of the Abelian Higgs model. It defines, that in the vacuum states
\begin{equation}
\label{vac1}  
\ph \Big|_{vac} =  1.  
\end{equation}  
Note however, that one does not need to impose this condition at spacial infinity ($r \rar \infty$) in order to get the finite action because of the factor $1/r^2$ in front of the corresponding potential term.    

The solution interpolating between different vacuum states is an instanton in the Abelian Higgs model. Interestingly (see \cite{Witten_inst, WulPom2}), the baryon number (\ref{baryon}) expressed in terms of 2D fields (\ref{ansatz})  
\begin{equation}  
\label{top_charge}  
Q_B = \frac{1}{2 \pi} \int dr \int \limits_\epsilon^{z_m} dz \ \epsilon_{ab} \Big(\p_a (-i \boldsymbol{\ph} D_b \overline{\boldsymbol{\ph}} - h.c.) + F_{ab} \Big)  
\end{equation}  
coincides with the topological charge of 2D instanton. Rewriting $\p_a (-i \boldsymbol{\ph} D_b \overline{\boldsymbol{\ph}} - h.c.) = 2\p_a (\ph^2 \p_b \alpha)$ we see, that the nonzero charge is related to the nontrivial behavior of the phase $\alpha$.      

The second potential term in the action (\ref{2D-action}) is the last one in the second line. It is convenient to rewrite it in terms of moduli and phases of the scalar fields:  
\begin{equation}  
\label{potential}  
\frac{\Lambda^2}{2 z^3} (|\boldsymbol{\chi}|^2 + |\boldsymbol{\ph}|^2 - |\boldsymbol{\ph} + \boldsymbol{\chi}|^2)  = \frac{\Lambda^2}{z^3} \chi^2 \ph^2 \ \cos(\alpha - \beta)^2  
\end{equation}  
We see, that the vacuum state is now defined by the relative phase of scalar fields. In what follows we will use the special notion for it  
\begin{equation}  
\gamma = \alpha - \beta - \frac{\pi}{2}.  
\end{equation}  
In the vacuum state one has  
\begin{equation}  
\label{2nd_charge}  
\gamma \big|_{vac} = \pi N, \quad N \in \mathbb{Z}.  
\end{equation}  
This is in fact the manifestation of the additional topological quantum number $N$, which counts these discrete vacua and appears in the theory, when the scalar field is included. The aim of this work is to describe the classical solution possessing an additional charge $Q_5 \neq 0$, related to this quantum number, which is presumably connected to the axial charge of the baryon. Note also, that the potential term under consideration is not suppressed at infinity, so in order to have finite action we require 
\begin{equation}
\label{boundary1}  
\chi^2 \ph^2 \ \sin(\gamma)^2 \Big|_{r \rar \infty} = 0.  
\end{equation}    

Following the usual ideology of constructing the instanton solution, we would like to have a field configuration, which interpolates between different vacua at different boundaries of the space. Our next step is to study by looking at the asymptotics of equations of motion whether the nontrivial behavior of the phases along the spacial boundary is possible. The equations of motion following from the action (\ref{2D-action}) are  
\begin{align}  
\ph:& & &\p_z \frac{1}{z} \p_z \ph + \frac{1}{z} \p_r^2 \ph = \frac{1}{z} \ph \Big[(\p_z \gamma + \p_z \beta)^2 + (\p_r \gamma + \p_r \beta + A_r)^2 \Big] \\
\notag  
& &   & \qquad  \qquad \qquad  \qquad \qquad - \frac{1}{r^2 z} \ph (1 - \ph^2) + 12 \frac{1}{z^3} \chi^2 \ph \sin(\gamma)^2, \\  
\chi: & & &\p_z \frac{r^2}{z^3} \p_z \chi + \p_r \frac{r^2}{z^3} \p_r \chi = \frac{r^2}{z^3} \chi \Big[(\p_z \beta)^2 + (\p_r \beta + A_r)^2 \Big] \\
\notag  
& & & \qquad  \qquad \qquad  \qquad \qquad + 8 \frac{1}{z^3} \chi \ph^2  \sin(\gamma)^2 - 3 \frac{r^2}{z^5} \chi, \\
\beta: & & &\p_z \frac{2}{z} \ph^2 (\p_z \gamma + \p_z \beta) + \p_r \frac{2}{z} \ph^2 (\p_r \gamma + \p_r \beta + A_r) \\  
\notag  
& & & \qquad  \qquad \qquad  \qquad \qquad + \p_z \frac{3 r^2}{z^3} \chi^2 \p_z \beta + \p_r \frac{3 r^2}{z^3} \chi^2 (\p_r \beta + A_r) =0,  \\  
\gamma: & & &\p_z \frac{1}{z} \ph^2 (\p_z \gamma + \p_z \beta) +  \p_r \frac{1}{z} \ph^2 (\p_r \gamma + \p_r \beta + A_r) = -\frac{6}{z^3} \chi^2 \ph^2 \ \sin(2\gamma),\\  
A_r: & &  &\p_z \frac{r^2}{z} \p_z A_r = \frac{4}{z} \ph^2 (\p_r \gamma + \p_r \beta + A_r) + \frac{6 r^2}{z^3} \chi^2 (\p_r \beta + A_r),  
\end{align}  
where we used phases $\beta$ and $\gamma$ as independent fields and fixed the gauge $A_z = 0$.    

\begin{figure}[h!]  
\centering    
\subfloat[][$z=0$]{
  \begin{minipage}[h]{0.24\linewidth}  	
  \includegraphics[width=1 \linewidth]{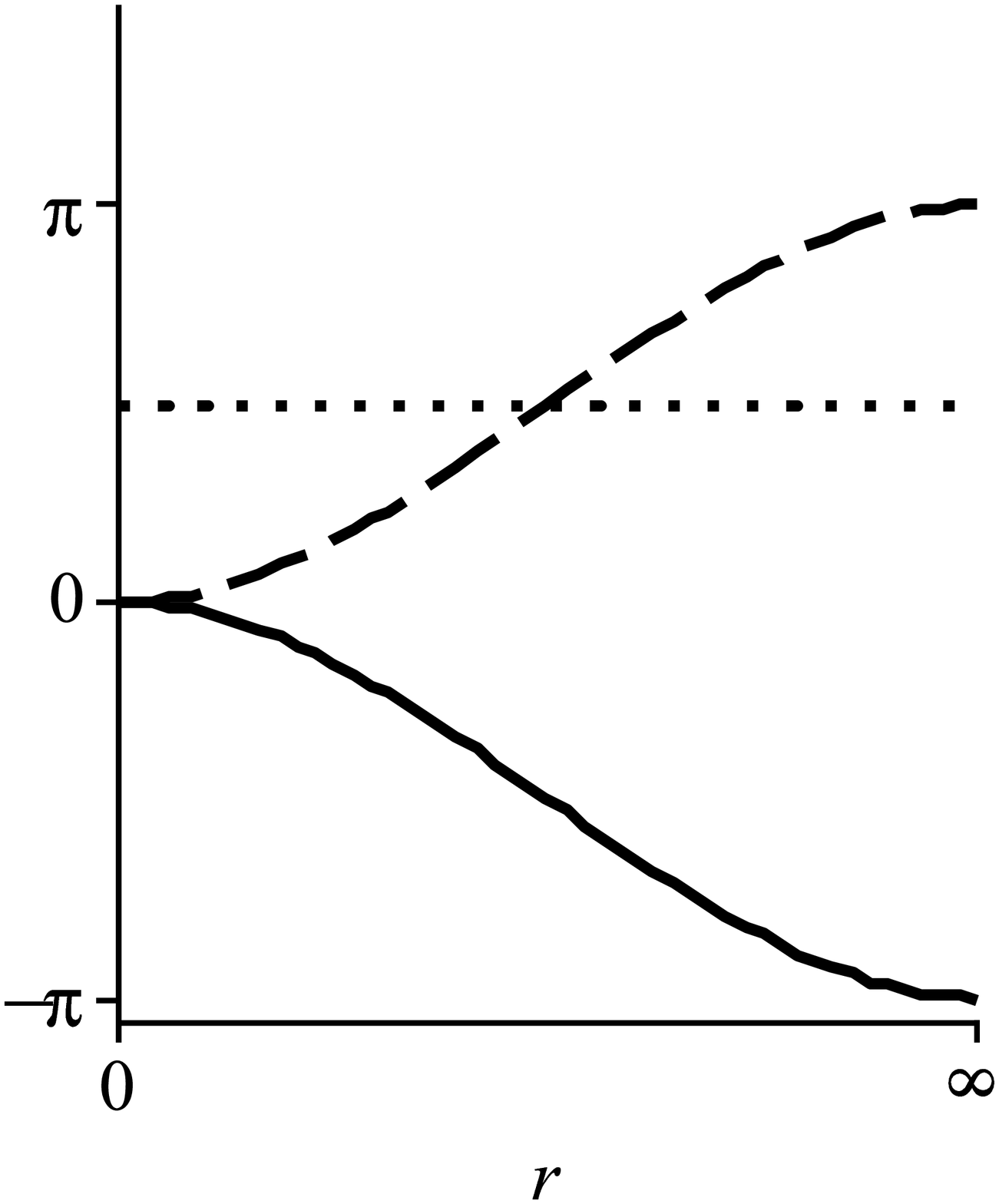}  	
  \end{minipage}}        
\subfloat[][$r \rar \infty$]{  	
  \begin{minipage}[h]{0.24\linewidth}  	
  \includegraphics[width=1 \linewidth]{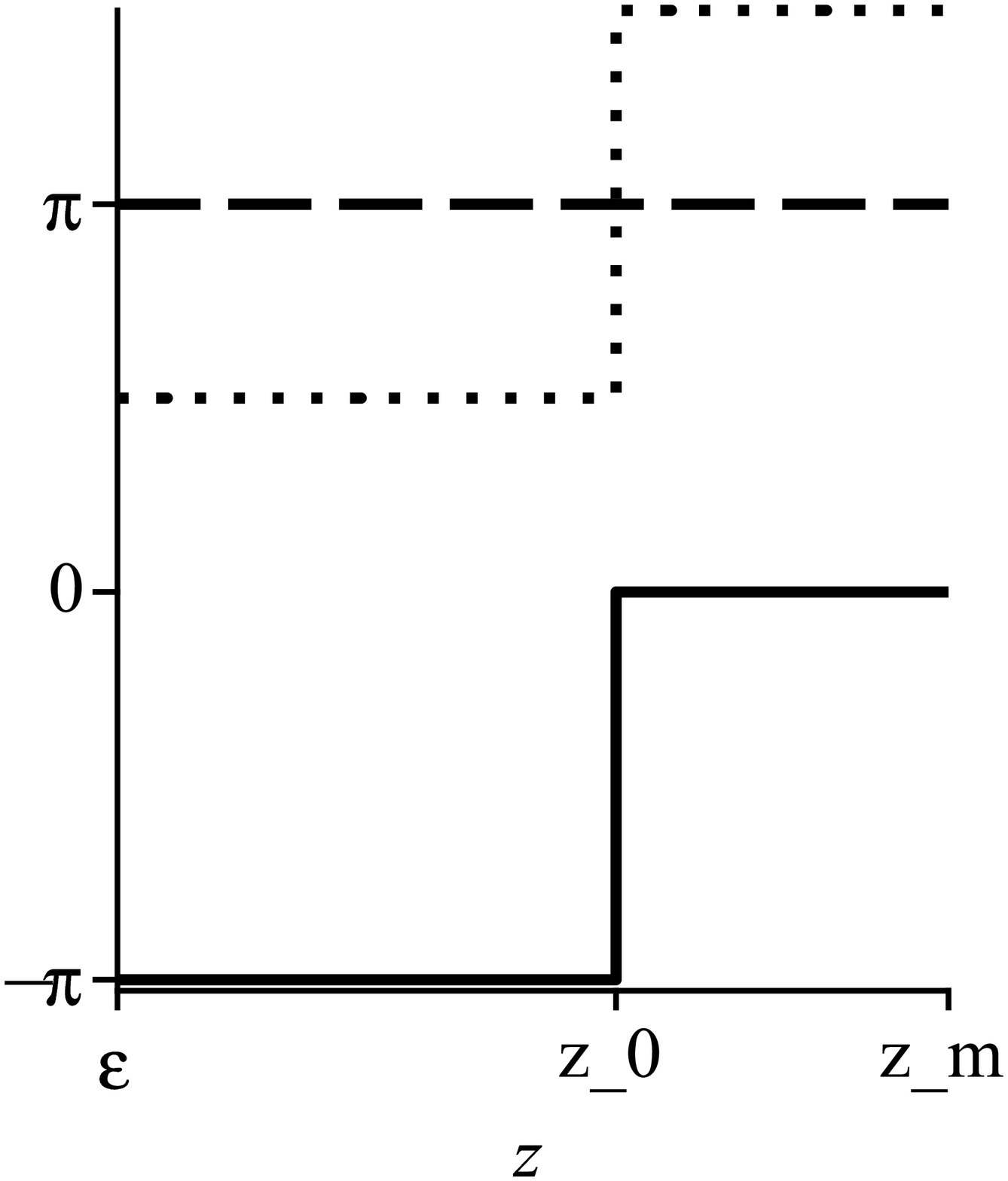}  	
  \end{minipage}}     
\subfloat[][$z=z_m$]{  	
  \begin{minipage}[h]{0.24\linewidth}  	
  \includegraphics[width=1 \linewidth]{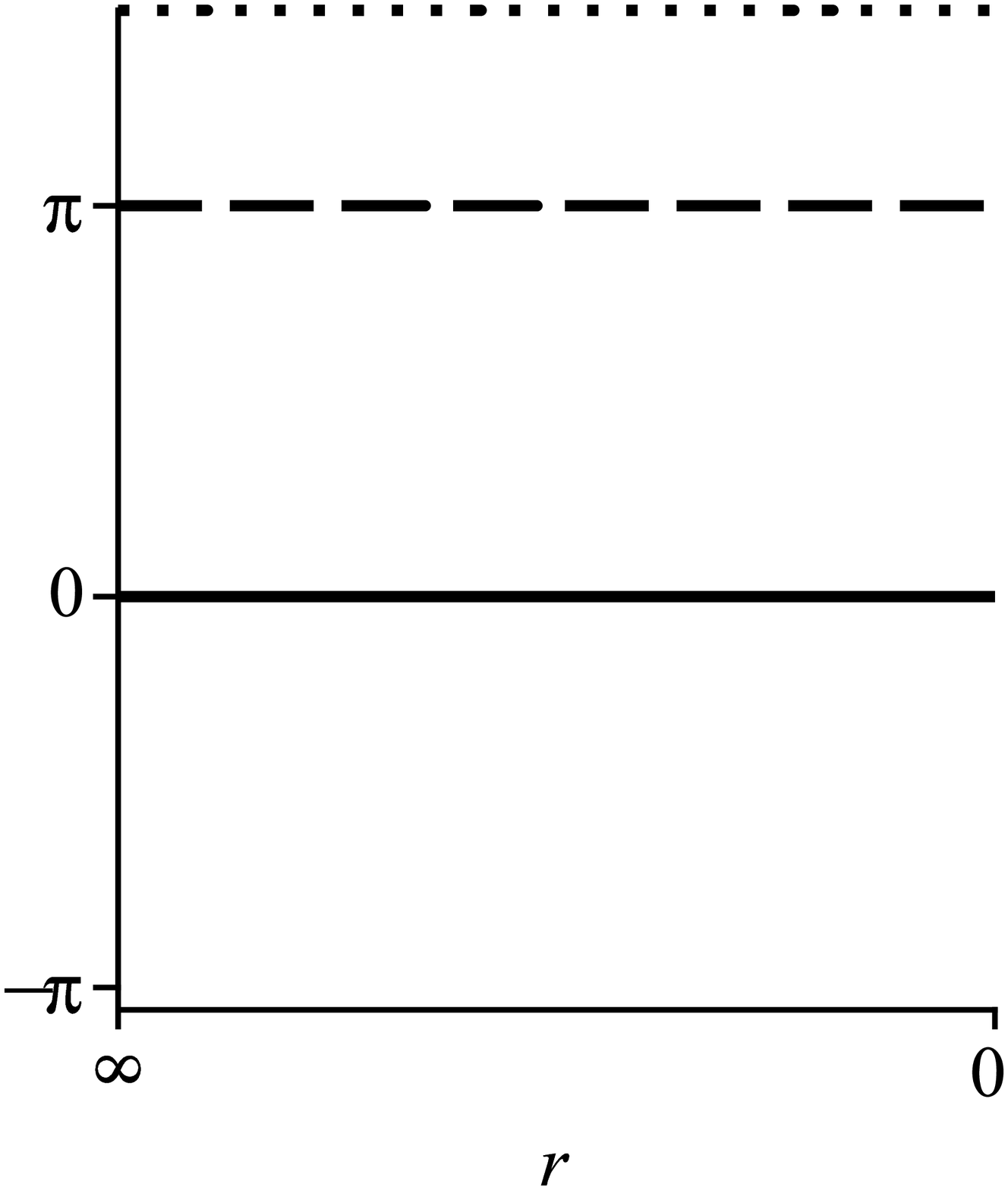}  	
  \end{minipage}}    
\subfloat[][$r=0$]{  	
  \begin{minipage}[h]{0.24\linewidth}  	
  \includegraphics[width=1 \linewidth]{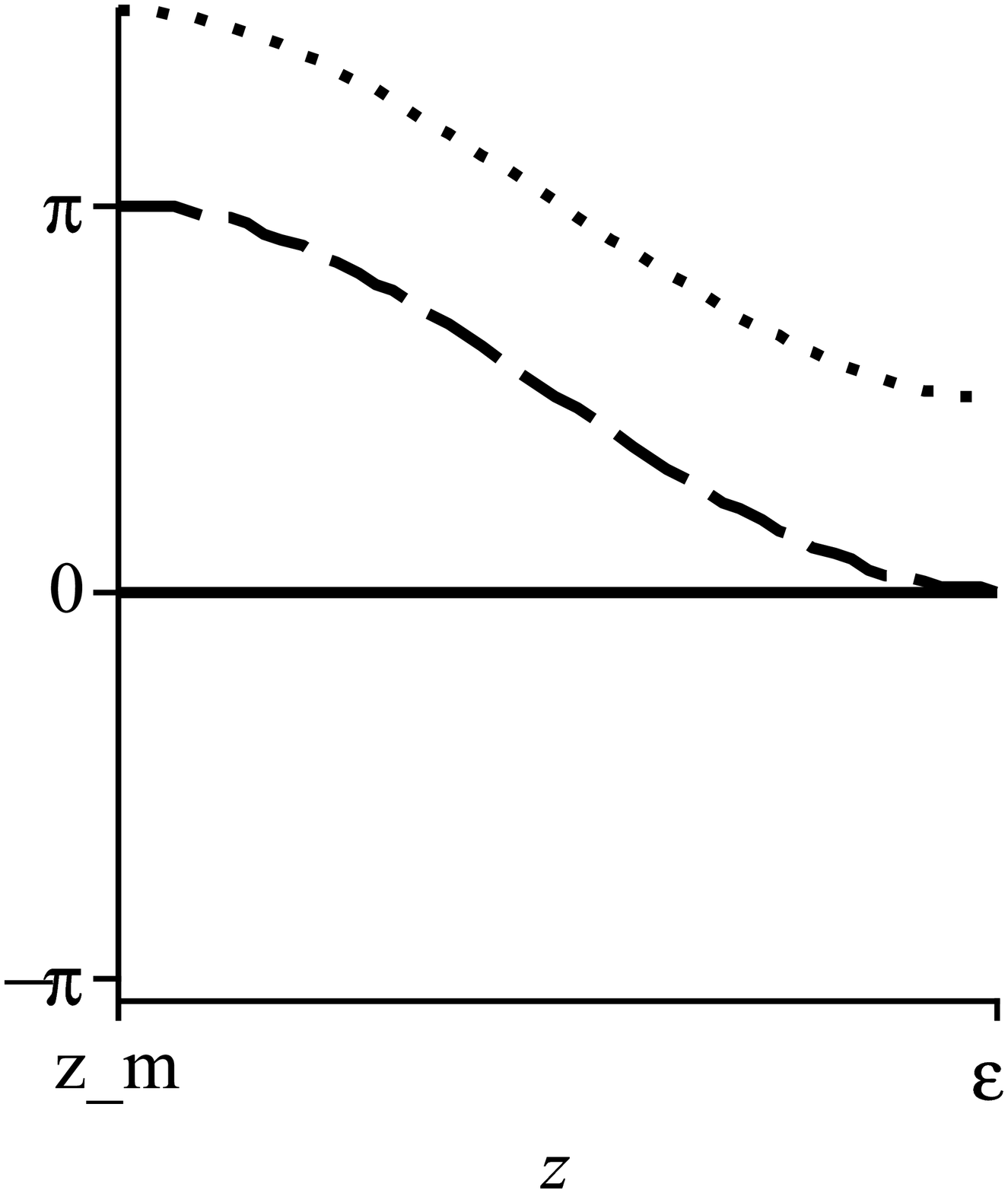}  	
  \end{minipage}}  
\caption{\label{a} The qualitative boundary behavior of the phases $\alpha$ (doted), $\beta$ (dashed) and $\gamma$ (solid). (Horizontal axis of (c) and (d) reversed.)  }  
\end{figure}

Let us start from the boundary $\{r=0, z \in (0,z_m)\}$. The requirement of the finiteness of the action (\ref{2D-action}) at this boundary leads to the condition  
\begin{equation}  
\ph(r,z) \Big|_{r=0} = 1.  
\end{equation}  
As the equations of motion are singular, we need to impose more requirements on boundary conditions in order to avoid singularities in the solutions. The straightforward way to figure out these requirements is to expand the expressions for the differential operators of the second order, appearing in the equations of motion, near the corresponding boundary. In other words, we require the second derivative, transverse to the boundary, to be finite on this boundary. At $r=0$ this gives:  
\begin{equation}  
\label{r0_boundary}  
\gamma(0,z) = 0, \qquad \p_r \phi(0,z) =0 , \qquad \p_r \chi(0,z) =0  
\end{equation}  
We see, that although $\gamma$ is fixed along the boundary, $\beta$ is unconstrained, so two phases $\alpha$ and $\beta$ can change simultaneously, keeping their difference constant (see fig.\ref{a}(d)). The change of $\alpha$ will contribute to the topological charge (\ref{top_charge}). One can check, that in order to get the smooth solution for the gauge field in ansatz (\ref{ansatz}) we need to require $\alpha = -\frac{\pi}{2}$, corresponding to $\ph_2=-1$, but as was pointed out in \cite{Witten_inst}, provided $\ph=1$ a singular gauge transformation can be found, which drives a singular solution to the smooth one.

The $\{r \in (0,\infty), z=0\}$ boundary can also be analyzed. In order to keep the action finite we have to demand  
\begin{align}  
\ph(r,z) \Big|_{z=0} = 1, \qquad \chi(r,z) \Big|_{z=0} \sim \sigma z^3, \qquad A_r(r,z) \Big|_{z=0} \sim z^2  
\end{align}  
Looking to the asymptotic behavior of the equations of motion, we see the following constraints
\begin{align}  
\label{z0_boundary}  
\p_r \alpha(r,\epsilon) \equiv \p_r \gamma(r,\epsilon) + \p_r \beta(r,\epsilon) &= -A_r(r,\epsilon) \sim \epsilon^2 \Big|_{\epsilon=0}, \\  
\p_z \alpha(r,\epsilon) \equiv \p_z \gamma(r,\epsilon) + \p_z \beta(r,\epsilon) &= 0. \notag  
\end{align}  
These conditions prohibit the shift of $\alpha$, but again leave two other phases free. We will consider a flow of $\gamma$ along the $z=0$ boundary from $0$ to $-\pi$ to get the solution with nontrivial second topological number (\ref{2nd_charge}). Accordingly, the phase $\beta$ will flow from $0$ to $\pi$ (see fig.\ref{a}(a)). Note, that this behavior of $X$ is just the same as expected for usual Skyrmion solution in the boundary theory, where the simplest pion field configuration is:  
\begin{equation}  
U(r) = \mathbf{1} \sin(\theta(r)) + \pi^a \tau^a \cos(\theta(r)),  
\end{equation}  
where $\theta(0) = 0$ and $\theta(\infty)=\pi$. We will discuss the connection between the running  of $\beta$ and the axial charge below.    

The hard wall boundary $\{r \in (0,\infty), z=z_m\}$ in our system is a bit special, because the boundary conditions on it are not dictated by the singular properties of action or equations of motion, but are imposed by hand according to the basic features of the hard-wall model (see \cite{Erlich, WulPom2}). In order to be in agreement with \cite{Erlich} while keeping in mind, that the ``hard-wall'' is not the only possible AdS/QCD model and one can consider the models with infinite IR radius $z_m \rar \infty$ \cite{soft_wall}, we impose the conditions at $z=z_m$, which forbid any dynamics and demand the action to vanish (see fig.\ref{a}(c)). We argue, that these boundary conditions will lead to the finite action in any AdS/QCD model.    
\begin{gather}  
\label{zm_boundary}  
\p_r \alpha(r,z_m) =  \p_r \beta(r,z_m) = \p_r \gamma(r,z_m) =0,\\  
\ph(r,z_m) = 1. 
\notag  
\end{gather}    

The only boundary left unexplored is the radial infinity $\{r \rar \infty, z \in (0,z_m)\}$. The finiteness of the action requires  
\begin{equation}  
\label{inf_boundary}  
\ph^2 \chi^2 \sin(\gamma)^2 \Big|_{r \rar \infty} = 0,  
\end{equation}  
but note, that we are not forced to impose $\ph=1$ condition here. This fact opens an interesting possibility to fulfill (\ref{inf_boundary}) keeping the nontrivial dynamics of $\gamma$.  
\begin{figure}[h!]  
\includegraphics[width=0.3 \linewidth]{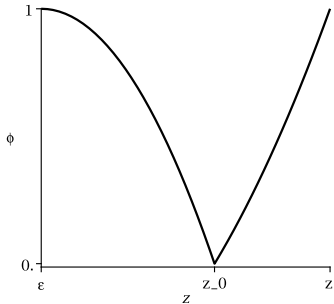}  
\caption{\label{b} Boundary value of the field $\ph$ at $r \rar \infty$. The field vanishes at $z=z_0$.}
\end{figure}  

On the boundary $\ph(r,z)$ must obey the equation of motion  
\begin{equation}  
\p_z^2 \ph(r,z) - \frac{1}{z} \ph(r,z) =0, \qquad r\rar \infty  
\end{equation}  
and we can consider the boundary value ($\theta$ is a Heaviside step function)  
\begin{align}  
\label{phi_inf}  
\ph(r,z) \Big|_{r \rar \infty} = &\theta(z_0-z) \left(1 - \frac{z^2}{z_0^2} \right) \\  
\notag   
+&\theta(z-z_0) \left( \frac{z^2 - z_0^2}{z_m^2 - z_0^2} \right)  
\end{align}  
which has proper values $\ph=1$ at the boundaries $z=0$ and $z=z_m$, but falls to zero at $z=z_0$ (see fig.\ref{b}). At this point the boundary value of $\gamma(r,z)$ can perform a jump from~$-\pi$~to~$0$ (see fig.\ref{a}(b))  
\begin{equation}  
\label{gamma_inf}  
\gamma(r,z)\Big|_{r \rar \infty} =-\pi + \theta(z-z_0) \ \pi,  
\end{equation}  
thus closing the contour of boundary values. While this arrangement seems to be singular, it doesn't lead to the divergence of the action. Indeed, in the kinetic term the derivative of $\alpha$, which diverges at $z=z_0$ in our solution, is multiplied by the modulus $\ph$, which is zero exactly in this point. Moreover, as  the behavior of $\ph$ is not smooth at $z=z_0$, its second derivative diverges, but in the action only first derivative of $\ph$ enters the kinetic term and this derivative is finite everywhere in the vicinity of $z_0$. We should point out also, that as the action is a two dimensional integral over $z$ and $r$, even a simple pole in the action density at point $\{z=z_0, r\rar \infty \}$ would not produce the divergence of the whole action.    

It should be underlined, that our solution differs substantially from a constant one, which was considered in \cite{Wulzer4} within the similar framework.  

We note, that the  solutions for $\alpha(z)$ and $\ph(z)$ respect the singular equations of motion at the boundary. The spectacular feature of these boundary values is that they represent a domain wall, located at $z=z_0$. The position of the wall governs mostly the effective mass of our instanton solution (see \cite{Sakai}) and we note, that the instanton can not ``fall'' on the hard wall, because of the boundary value of the field $\ph(r,z)$ at $z=z_m$. That means that the potential (\ref{potential}) at spatial infinity ($r \rar \infty$) stabilizes the mass and the size of the baryon.  We stress, that one do not need to consider any additional mechanisms of the stabilization of instanton size \cite{WulPom1,WulPom2,Sakai}, as it is fixed by the presence of the second topological charge of the solution itself.  This effect is somewhat inherited from the feature of the original dyonic instanton \cite{lt} whose size is stabilized by its quantum numbers (\ref{rho}).    

At the end of the day we obtained an interesting picture of asymptotics of our ``dyonic instanton'' (\ref{r0_boundary},\ref{z0_boundary},\ref{zm_boundary},\ref{phi_inf},\ref{gamma_inf}). The phase $\gamma$ is 0 at the $r=0$ boundary, then it flows to $-\pi$ along $z=0$ and jumps back to zero on $r\rar \infty$, thus interpolating between different vacua (\ref{2nd_charge}), realized at $r=0$ and point $r \rar \infty, z=0$. The phase $\alpha$ flows from $\tfrac{3\pi}{2}$ to $\tfrac{\pi}{2}$ along $r=0$, than stays constant at $z=0$ and jumps with $\gamma$ back to $\tfrac{3\pi}{2}$ on $r\rar \infty$. It realizes the nontrivial topological configuration with nonzero charge (\ref{top_charge}), showing that our solution is indeed a baryon.    

A few words are in order concerning the phase $\beta$. In the holographic action (\ref{action}) the axial field couples to the field $X$ via the kinetic term (see also \cite{Krik2}). Consequently the $X$ field can be a source to the axial current (dual to the axial field), namely  
\begin{equation}  
J^A_\mu \sim i (\p_\mu X X^\dag - X \p_\mu X^\dag).  
\end{equation}  
In the 2D fields notation, this expression for non-singlet current looks as  
\begin{equation}  
\label{axial_current}  
J^A_r \sim 2 \chi^2 \ \p_r \beta(r,z) \ \boldsymbol{\tau}.  
\end{equation}  
That means, that the phase $\beta$ governs the coupling of our solution with external axial current, and thus the axial charge of the baryon (\ref{axial_charge}).    

\section{On the brane picture}    

In this Section we shall make  few comments concerning the brane interpretation  of the solution under consideration.  The dyonic instantons can be considered  as the instantons (D0 branes) with attached fundamental strings   \cite{lt, town,dyonic2}.  More precisely the D0 branes are localized at the D4 branes  and fundamental string connects two parallel  D4 branes with the geometry $R^4\times I$ where $D=5$ gauge theory lives on.  The dyonic instanton size is fixed by the asymptotic distance between D4 branes.            

Let us mention the similarities between the "color" dyonic instanton \cite{lt, town}  and the "flavor" solution we have considered.   It is useful to have in mind the Sakai-Sugimoto type  geometry \cite{ss}, where the 5D gauge theory  is defined at the worldvolume of $N_f$ D8-$\bar{D8}$ branes  wrapped around the internal $S^4$ and extended along the radial coordinate  on the cigar. The branes and antibranes are localized at the different  points in the periodic coordinate on the cigar. Baryons are identified with the  D4 branes wrapped around $S^4$ \cite{Sakai, witten2}.     

To keep close to the standard dyonic instanton \cite{dyonic2} we assume that the brane  configuration involves also the fundamental string attached to the  D4 instanton and connecting left and right stacks of  $D8$ branes in the Sakai-Sugimoto framework \cite{Aharoni, berg}.  In our case similarly the VEV of the bifundamental scalar identified with the chiral   condensate is related  to the distance between the stacks like in \cite{berg}.        

More general dyonic instanton solution corresponds to  the (D0+ F1) state   blown  into the tubular D2 branes with the electric field  and the  instanton charge \cite{town}. That is in the $D=4$ space-time  the point-like instanton becomes the circular monopole  loop which carries the topological charge due to D0  density  and electric field due to the fundamental string.  Our spherically symmetric Skyrmion discussed above   is interpreted as a particle   representing the baryon.  However in more general solution we could expect that D4 brane  gets blown into the D6 brane \cite{town, dyonic2}. If such  D6 brane is extended in one space coordinate   the  solution looks like a closed string similar to the  monopole loop in the original Lambert-Tong case \cite{town}. The total magnetic charge   of the closed string vanishes but its  dipole magnetic moment survives. It would  be very interesting to compare the natural loop structure of  such more general  dyonic instanton solution with the loop structure observed numerically   for the Skyrmions with baryonic charge $B\geq 2$ \cite{torus} long time ago.        

We have mentioned above that in some range of parameters there is negative  mode around the dyonic  instanton configuration. The mode analysis in $AdS_4$  case has to be performed and we restrict ourselves by one remark.  It is known that monopole induces the baryon decay via Rubakov-Callan effect \cite{Rubakov-Callan}.  It has been realized holographically in  \cite{hong} where the process   has been described as the blow up of D4 brane into D6 brane somewhat similar  to  our analysis. We hope to discuss these issues elsewhere.                   

\section{Conclusion}  

In this paper we have discussed  the realization of the dyonic instanton solution within the framework of holographic QCD as baryon. We have  showed, that considering the   holographic model with bifundamental scalar field, dual to the chiral condensate of quarks, one can figure out two different topological numbers of the solution, corresponding to the baryonic and the axial charges of QCD baryon.    

The solution, discussed in this work has a peculiar form of domain wall, whose thickness, while being zero at radial infinity, rises closer to the core, covering the whole distance between holographic boundaries in the center of instanton. First of all this shape reminds the connection between Skyrmions and instantons inside a  domain wall, pointed out by Atiyah and Manton and developed by Tong at al. \cite{tongjap, am}. Second, the domain wall is forced to be located between two holographic boundaries, resolving the known problem of holographic instantons ``falling'' on a hard wall. Indeed, we do no need to consider special dynamics on the IR brane \cite{WulPom1} or the back reaction of the vector Abelian field, sourced by $Q_B$ on the solution \cite{WulPom2}, which turns out to be suppressed in certain models by the string scale (see discussion in \cite{WulPom2} and \cite{Sakai}). Instead, our solution by the construction has finite size, which is determined solely by the dynamics. The position of the domain wall, related to the mass scale of the baryon is  governed by the value of the chiral condensate $\sigma$, being the only tunable parameter of the model. That means, that our approach of describing baryons in holographic models provides a   solid ground  for the Ioffe relation (\ref{Ioffe}), which origin was otherwise rather mysterious.    

While the qualitative study of some features of the holographic dyonic instanton can be based on the consideration of the asymptotic  behavior of solution, obtaining of the full solution to the equations of motion with above mentioned asymptotics is highly desirable and we leave it for future work. The other interesting developments in this direction could be the consideration of the time components of gauge fields, giving rise to 4D electrical field strength and study of possible solutions, absent in the cylindrical anzatz, used here.   It would be very interesting to clarify the issue of  an angular momentum and its role in the stabilization  of the configuration. The possible magnetic dipole nature of the solution deserves the additional  investigation together with the possible negative modes at large quantum numbers.      

The separate issue concerns the temperature dependence of the solution  and its behavior under the chiral phase transition.  If there would be  nontrivial "holonomy" along the radial $AdS_4$ coordinate  one could expect caloron-like solution involving   "dressed monopoles" in the flavor group  with the fractional baryon  numbers. Such states are familiar in the theories  with the compact dimensions. Certainly all these issues deserve further investigation.    

\acknowledgments  

A.G. thanks M. Voloshin for the useful discussions and FTPI at University of Minnesota where a  part of the work has been done for the hospitality and support.  A.K. would like to thank David Tong for insightful comments and Isaac Newton Institute for Mathematical Sciences where a part of this work has been done. We  are  grateful to Pavel Khromov for participation at the early stage of the project. The research of A.K. is supported in part by RFBR grant no.12-02-00284 and PICS- 12-02-91052, the Ministry of Education and Science of the Russian Federation under contract 14.740.11.0347, and the Dynasty foundation. The work of A.G. is  supported in part by grants RFBR-12-02-00284 and PICS- 12-02-91052.


\begin{thebibliography}{99}  

\bibitem{witten1}    
E.~Witten,    
``Baryons in the 1/n Expansion,''    
Nucl.\ Phys.\ B {\bf 160}, 57 (1979).    

\bibitem{witten2}   
E.~Witten,    
``Baryons and branes in anti-de Sitter space,''    
JHEP {\bf 9807}, 006 (1998)    
[hep-th/9805112].    
%%CITATION = HEP-TH/9805112;%%    

\bibitem{son}   
D.~T.~Son and M.~A.~Stephanov,    
``QCD and dimensional deconstruction,''    
Phys.\ Rev.\ D {\bf 69}, 065020 (2004)    
[hep-ph/0304182].    
%%CITATION = HEP-PH/0304182;%%    

\bibitem{am}  
M.~F.~Atiyah and N.~S.~Manton,    
``Skyrmions From Instantons,''    
Phys.\ Lett.\ B {\bf 222}, 438 (1989) \\   
M.~Atiyah and P.~Sutcliffe,    
``Skyrmions, instantons, mass and curvature,''    
Phys.\ Lett.\ B {\bf 605}, 106 (2005)    
[hep-th/0411052].    
%%CITATION = HEP-TH/0411052;%%    

\bibitem{tongjap}   
M.~Eto, M.~Nitta, K.~Ohashi and D.~Tong,    
``Skyrmions from instantons inside domain walls,''    
Phys.\ Rev.\ Lett.\  {\bf 95}, 252003 (2005)    
[hep-th/0508130].    
%%CITATION = HEP-TH/0508130;%%      

\bibitem{Sakai}    
H.~Hata, T.~Sakai, S.~Sugimoto and S.~Yamato,    
``Baryons from instantons in holographic QCD,''    
Prog.\ Theor.\ Phys.\  {\bf 117}, 1157 (2007)    
[hep-th/0701280 [HEP-TH]].    
%%CITATION = HEP-TH/0701280;%%    

%\cite{Pomarol:2009hp}  
\bibitem{WulPom3}    
A.~Pomarol and A.~Wulzer,    
``Baryon physics in a five-dimensional model of hadrons,''    
arXiv:0904.2272 [hep-ph].    
%%CITATION = ARXIV:0904.2272;%%    

%\cite{Pomarol:2007kr}  
\bibitem{WulPom1}    
A.~Pomarol and A.~Wulzer,    
``Stable skyrmions from extra dimensions,''    
JHEP {\bf 0803}, 051 (2008)    
[arXiv:0712.3276 [hep-th]].    
%%CITATION = ARXIV:0712.3276;%%    

%\cite{Pomarol:2008aa}  
\bibitem{WulPom2}    
A.~Pomarol and A.~Wulzer,    
``Baryon Physics in Holographic QCD,''    
Nucl.\ Phys.\ B {\bf 809}, 347 (2009)    
[arXiv:0807.0316 [hep-ph]].    
%%CITATION = ARXIV:0807.0316;%%    

\bibitem{fermi_baryon}    
T.~Gutsche, V.~E.~Lyubovitskij, I.~Schmidt and A.~Vega,      
``Nucleon structure including high Fock states in AdS/QCD,''  
Phys.\ Rev.\ D {\bf 86}, 036007 (2012)  
[arXiv:1204.6612 [hep-ph]].  \\    
%%CITATION = ARXIV:1204.6612;%%     
H.~Forkel, M.~Beyer and T.~Frederico,      
``Linear meson and baryon trajectories in AdS/QCD,''  
Int.\ J.\ Mod.\ Phys.\ E {\bf 16}, 2794 (2007)  
[arXiv:0705.4115 [hep-ph]].  \\    
%%CITATION = ARXIV:0705.4115;%%    
S.~J.~Brodsky and G.~F.~de Teramond,      
``Hadronic spectra and light-front wavefunctions in holographic QCD,''  
Phys.\ Rev.\ Lett.\  {\bf 96}, 201601 (2006)  
[hep-ph/0602252].  \\    
%%CITATION = HEP-PH/0602252;%%    
D.~K.~Hong, T.~Inami and H.~-U.~Yee,      
``Baryons in AdS/QCD,''  
Phys.\ Lett.\ B {\bf 646}, 165 (2007)  
[hep-ph/0609270].     
%%CITATION = HEP-PH/0609270;%%      

\bibitem{ioffe}    
B.~L.~Ioffe,    
``Calculation of Baryon Masses in Quantum Chromodynamics,''    
Nucl.\ Phys.\ B {\bf 188}, 317 (1981)    [Erratum-ibid.\ B {\bf 191}, 591 (1981)].    
%%CITATION = NUPHA,B188,317;%%    

\bibitem{lt}     
N.~D.~Lambert and D.~Tong,    
``Dyonic instantons in five-dimensional gauge theories,''    
Phys.\ Lett.\  B {\bf 462}, 89 (1999)    
[arXiv:hep-th/9907014].    
%%CITATION = PHLTA,B462,89;%%    

\bibitem{cs}    
S.~Kim, K.~-M.~Lee and S.~Lee,    
``Dyonic Instantons in 5-dim Yang-Mills Chern-Simons Theories,''    
JHEP {\bf 0808}, 064 (2008)    
[arXiv:0804.1207 [hep-th]].\\    
%%CITATION = ARXIV:0804.1207;%%     
B.~Collie and D.~Tong,    
``Instantons, Fermions and Chern-Simons Terms,''    
JHEP {\bf 0807}, 015 (2008)    
[arXiv:0804.1772 [hep-th]].    
%%CITATION = ARXIV:0804.1772;%%    
\bibitem{Erlich}    
Joshua Erlich, Emanuel Katz, Dam T. Son, Mikhail A. Stephanov,   
``QCD and a holographic model of hadrons'', 
SLAC-PUB-10965,    WM-05-101, INT-PUB-05-02, {\it Phys. Rev. Lett.} {\bf 95}, 261602,    2005

\bibitem{Witten_inst}    
E.~Witten,   
``Some Exact Multi - Instanton Solutions of Classical Yang-Mills Theory,''
Phys.\ Rev.\ Lett.\  {\bf 38}, 121 (1977).    
%%CITATION = PRLTA,38,121;%%    
\bibitem{dyonic2}    
D.~S.~Bak and K.~M.~Lee,    
``Supertubes connecting D4 branes,''    
Phys.\ Lett.\  B {\bf 544}, 329 (2002)    
[arXiv:hep-th/0206185].\\    
%%CITATION = PHLTA,B544,329;%%     
H.~Y.~Chen, M.~Eto and K.~Hashimoto,    
``The shape of instantons: Cross-section of supertubes and dyonic    instantons,''    
JHEP {\bf 0701}, 017 (2007)    
[arXiv:hep-th/0609142].    
%%CITATION = JHEPA,0701,017;%%    

\bibitem{town}    D.~Mateos and P.~K.~Townsend,    
``Supertubes,''    
Phys.\ Rev.\ Lett.\  {\bf 87}, 011602 (2001)    
[arXiv:hep-th/0103030].\\    
%%CITATION = PRLTA,87,011602;%%    
E.~Eyras, P.~K.~Townsend and M.~Zamaklar,    
``The heterotic dyonic instanton,''    
JHEP {\bf 0105}, 046 (2001)    
[arXiv:hep-th/0012016].\\    
%%CITATION = JHEPA,0105,046;%%    
D.~Mateos, S.~Ng and P.~K.~Townsend,    
``Tachyons, supertubes and brane/anti-brane systems,''    
JHEP {\bf 0203}, 016 (2002)    
[arXiv:hep-th/0112054].\\    
%%CITATION = JHEPA,0203,016;%%    
P.~K.~Townsend,    
``Surprises with angular momentum,''    
Annales Henri Poincare {\bf 4}, S183 (2003)    
[arXiv:hep-th/0211008].    
%%CITATION = AHPJF,4,S183;%%    

\bibitem{emparan}   
R.~Emparan, D.~Mateos and P.~K.~Townsend,    
``Supergravity supertubes,''    
JHEP {\bf 0107}, 011 (2001)    
[arXiv:hep-th/0106012].    

\bibitem{Krik1} 
 A.~Krikun,    
``On two-point correlation functions in AdS/QCD,''    
Phys.\ Rev.\  D {\bf 77}, 126014 (2008)    [arXiv:0801.4215 [hep-th]].    
%%CITATION = PHRVA,D77,126014;%%    

\bibitem{Krik2}    
A.~Gorsky and A.~Krikun,    
``Magnetic susceptibility of the quark condensate via holography,''   
Phys.\ Rev.\  D {\bf 79}, 086015 (2009)    
[arXiv:0902.1832 [hep-ph]].    
%%CITATION = PHRVA,D79,086015;%%    

\bibitem{Wulzer4}    
O.~Domenech, G.~Panico and A.~Wulzer,    
``Massive Pions, Anomalies and Baryons in Holographic QCD,''    
Nucl.\ Phys.\ A {\bf 853}, 97 (2011)    [arXiv:1009.0711 [hep-ph]].    
%%CITATION = ARXIV:1009.0711;%%    

\bibitem{soft_wall}    
A.~Karch, E.~Katz, D.~T.~Son and M.~A.~Stephanov,    
``Linear Confinement and AdS/QCD,''    
Phys.\ Rev.\  D {\bf 74}, 015005 (2006)    
[arXiv:hep-ph/0602229].    
%%CITATION = PHRVA,D74,015005;%%    

\bibitem{ss}    T.~Sakai, S.~Sugimoto,    
``Low energy hadron physics in holographic QCD,''    
Prog.\ Theor.\ Phys.\  {\bf 113}, 843-882 (2005).    
[hep-th/0412141].    

\bibitem{berg}    
O.~Bergman, S.~Seki and J.~Sonnenschein,    
``Quark mass and condensate in HQCD,''    
JHEP {\bf 0712}, 037 (2007)    
[arXiv:0708.2839 [hep-th]].    
%%CITATION = ARXIV:0708.2839;%%    

\bibitem{Aharoni}    
O.~Aharony and D.~Kutasov,      
``Holographic Duals of Long Open Strings,''  
Phys.\ Rev.\ D {\bf 78}, 026005 (2008)  
[arXiv:0803.3547 [hep-th]].    
%%CITATION = ARXIV:0803.3547;%%    

\bibitem{torus}    
V.~B.~Kopeliovich and B.~E.~Stern,    
``Exotic Skyrmions,''    
JETP Lett.\  {\bf 45}, 203 (1987)    
[Pisma Zh.\ Eksp.\ Teor.\ Fiz.\  {\bf 45}, 165 (1987)].\\    
%%CITATION = JTPLA,45,203;%%   
J.~J.~M.~Verbaarschot,    
``Axial Symmetry Of Bound Baryon Number Two Solution Of The Skyrme Model,''    
Phys.\ Lett.\ B {\bf 195}, 235 (1987).    
%%CITATION = PHLTA,B195,235;%%    

\bibitem{Rubakov-Callan}    
V.~A.~Rubakov,    
``Superheavy Magnetic Monopoles and Proton Decay,''     
JETP Lett.\  {\bf 33} (1981) 644   [Pisma Zh.\ Eksp.\ Teor.\ Fiz.\  {\bf 33} (1981) 658]. \\    
%%CITATION = JTPLA,33,644;%%     
V.~A.~Rubakov,    
``Adler-Bell-Jackiw Anomaly and Fermion Number Breaking in the Presence of a Magnetic Monopole,''     
Nucl.\ Phys.\ B {\bf 203}, 311 (1982).  \\
%%CITATION = NUPHA,B203,311;%%    
C.~G.~Callan, Jr.,    
``Monopole Catalysis of Baryon Decay,''  
Nucl.\ Phys.\ B {\bf 212}, 391 (1983).  \\
%%CITATION = NUPHA,B212,391;%%      
C.~G.~Callan, Jr. and E.~Witten,    
``Monopole Catalysis Of Skyrmion Decay,''  
Nucl.\ Phys.\ B {\bf 239}, 161 (1984).   \\  
%%CITATION = NUPHA,B239,161;%%    

\bibitem{hong}    
D.~K.~Hong, K.~-M.~Lee, C.~Park and H.~-U.~Yee,    
``Holographic Monopole Catalysis of Baryon Decay,''    
JHEP {\bf 0808}, 018 (2008)    
[arXiv:0804.1326 [hep-th]].    
\end{thebibliography}
\end{document}